\documentclass[%
    aps,
    prd,
    10pt,
    nofootinbib,
    noeprint,
    amsmath,
    amssymb,
    superscriptaddress
]{revtex4-2}

\usepackage{graphicx}
\usepackage[%
    colorlinks=true,
    allcolors=blue
]{hyperref}
\usepackage[%
    print-unity-mantissa=false,
    range-phrase=-,
    range-units=single
]{siunitx}

\def\araa{Annu.\ Rev.\ Astron.\ Astrophys.}
\def\apjs{Astrophys.\ J.\ Suppl.\ Ser.}
\def\cqg{Classical\ Quant.\ Grav.}
\def\cmp{Commun.\ Math.\ Phys.}
\def\cse{Comput.\ Sci.\ Eng.}
\def\epja{Eur.\ Phys.\ J.\ A}
\def\lrr{Living\ Rev.\ Relativity}
\def\mnras{Mon.\ Not.\ R.\ Astron.\ Soc.}

\def\pr{Phys.\ Rev.}

\def\rmp{Rev.\ Mod.\ Phys.}

\newcommand{\m}{{\sf m}}

\begin{document}

\title{Neutron-star seismology with realistic, finite-temperature nuclear matter}

\author{Fabian Gittins}
\email{f.w.r.gittins@uu.nl}
\affiliation{Institute for Gravitational and Subatomic Physics (GRASP), Utrecht University, Princetonplein 1, 3584 CC Utrecht, Netherlands}
\affiliation{Nikhef, Science Park 105, 1098 XG Amsterdam, Netherlands}
\affiliation{Mathematical Sciences and STAG Research Centre, University of Southampton, Southampton SO17 1BJ, United Kingdom}

\author{Nils Andersson}
\email{n.a.andersson@soton.ac.uk}
\affiliation{Mathematical Sciences and STAG Research Centre, University of Southampton, Southampton SO17 1BJ, United Kingdom}

\date{\today}

\begin{abstract}
    The oscillation spectrum of a neutron star is notably rich and intrinsically dependent on the equation of state of nuclear matter. With recent advancements in gravitational-wave and electromagnetic astronomy, we are nearing the capability to perform neutron-star asteroseismology and probe the complex physics of neutron stars. With this in mind, we explore the implementation of three-parameter finite-temperature matter models in the computation of neutron-star oscillations. We consider in detail the thermodynamics of nuclear matter and show how this information enters the problem. Our realistic treatment takes into account entropy and composition gradients that exist in the nuclear matter, giving rise to buoyant \textit{g}-mode oscillations. To illustrate the implementation, we determine the oscillation spectrum of a low-temperature neutron star. In addition to the expected compositional and thermal \textit{g}-modes, we find perturbations sourced by phase transitions in the equation of state. We also examine two thermal models, comparing the results for constant redshifted temperature with those for uniform entropy per baryon.
\end{abstract}

\maketitle

\section{Introduction}

Neutron stars are among the most complex objects in our Universe. A realistic description of a neutron star inevitably requires physics across the extremes; from Einstein's general relativity---to describe the immense gravitational fields---to quantum chromodynamics---in order to characterise the strong nuclear interactions in the dense interiors. The situation is no less complicated by the cataclysmic events that neutron stars are involved in, such as binary mergers.
Indeed, the celebrated event GW170817 \cite{2017PhRvL.119p1101A} marked the first gravitational-wave observation of a neutron-star coalescence and among the main science goals with these detections is to constrain the properties of ultra-dense nuclear matter \cite{2018PhRvL.121p1101A,2019PhRvX...9a1001A}. This effort will be further bolstered by the construction of the next generation of ground-based instruments, Cosmic Explorer \cite{2019BAAS...51g..35R} and the Einstein Telescope \cite{2010CQGra..27s4002P}.

The full complexity of neutron stars is arguably best evidenced by their remarkable oscillation spectra \cite{1988ApJ...325..725M,2015PhRvD..92f3009K}. Neutron stars possess oscillations that are linked to the pressure (acoustic \textit{f}- and \textit{p}-modes), thermal and compositional stratification (buoyant \textit{g}-modes), phase transitions (interfacial \textit{i}-modes), the presence of an elastic crust (shear \textit{s}-modes) and so on. As a guiding principle, each ingredient of physics that is added to a stellar model gives rise to a new class of oscillation mode.

The vibrational modes of neutron stars may be relevant in a variety of physical scenarios. There currently exists a promising connection between theory and measured quasi-periodic oscillations in the X-ray tails of giant magnetar flares \cite{2006ApJ...637L.117W,2007MNRAS.374..256S}. Along a similar vein, about 20 low-mass X-ray binaries have exhibited thermonuclear burst oscillations at frequencies up to \qty{3}{\hertz} different to the stellar rotation \cite{2012ARA&A..50..609W,bursts}, which have been conjectured to arise from stellar pulsations \cite{1996ApJ...460..827B,2004ApJ...600..939H}. (However, the association may be problematic as oscillation frequencies identical to the neutron star's spin cannot have a mode origin \cite{2023Univ....9..226G}.) To complement these electromagnetic detections, neutron-star modes may leave their mark on gravitational-wave astronomy \cite{1998MNRAS.299.1059A}. The gravitational-wave instability associated with the \textit{r}-mode \cite{1998ApJ...502..708A,1998ApJ...502..714F} has lead to recent searches \cite{2021ApJ...922...71A} and oscillation modes will also be involved with the tidal response in compact-binary inspirals \cite{1994MNRAS.270..611L}, where the predominant contribution  comes from the \textit{f}-mode and low-frequency oscillations may be excited at a lower level. Conducting neutron-star asteroseismology with these observational channels has the exciting potential to constrain the rich physics associated with the modes, including the dense nuclear-matter equation of state.

In this work, we compute the oscillation spectra of neutron stars described by realistic, finite-temperature nuclear matter. Our treatment moves beyond phenomenological thermal-pressure models that are often used in the literature (see, \textit{e.g.}, Refs.~\cite{2015PhRvD..92f3009K,2022Galax..10...79L}) by implementing a finite-temperature equation-of-state table. By providing this information at the level of the equation of state, we model realistic stratification due to entropy and composition gradients in the matter, giving rise to gravity \textit{g}-mode oscillations. Additionally, our assumed matter model---the APR equation of state \cite{2019PhRvC.100b5803S}---involves a phase transition. We identify the corresponding interface \textit{i}-mode and explore its dependence on stratification. Our inclusion of finite temperature is a development with respect to recent work on neutron-star \textit{g}-modes at zero temperature \cite{2021PhRvD.103l3009J,2021PhRvD.104l3032C,2022PhRvD.105j3025Z,2025MNRAS.536.1967C} and our numerical implementation was used to support a study on the thermal effects in merger simulations \cite{2025PhRvD.111b3049G}.

For completeness, we carefully develop the machinery required to incorporate realistic, finite-temperature nuclear-matter models into calculations of neutron-star oscillations. We do this to address two issues that may not be as well understood as they should. The first relates to what the thermodynamical equation of state of nuclear matter actually is. It is common in the neutron-star community to refer to the function $\varepsilon = \varepsilon(p)$ as the equation of state. However, for reasons we lay out in this paper, such a relation will never provide the level of detail required to appropriately use neutron-star observations to probe the behaviour of dense nuclear matter. The second reason follows from the fact that the bulk of the astrophysical literature until now has focused on simple, barotropic models for the stellar material. As we discuss here, this treatment of the matter will only apply for equilibrium neutron stars. For dynamical contexts, we must consider departures from the barotropic description.

There are notable exceptions to the trend of assuming one-parameter matter models. This includes work on incorporating realistic stratification with \cite{2021PhRvD.103l3009J,2021PhRvD.104l3032C,2025MNRAS.536.1967C} and without the Cowling approximation \cite{2022PhRvD.105j3025Z}, superfluidity \cite{2001MNRAS.328.1129A,2002PhRvD..66j4002A,2004MNRAS.348..625P,2008PhRvD..78h3008L} and crustal physics \cite{2015PhRvD..92f3009K,2019CQGra..36j5004A} into the modelling. There is also a large body of literature dedicated to understanding the oscillation spectra of compact remnants following core-collapse supernovae (\textit{e.g.}, Refs.~\cite{2003MNRAS.342..629F,2018MNRAS.474.5272T,2018ApJ...861...10M,2019MNRAS.482.3967T,2023MNRAS.523.2236R,2023PhRvL.131s1201J}). In this context, the matter is hot and out of equilibrium, and the oscillation studies rely on data extracted from general-relativistic hydrodynamical simulations. The latter point is a notable difference with respect to the numerical implementation that we adopt in this paper, where we solve for both the background and the fluid perturbations in a coupled fashion. Furthermore, we go beyond the state of the art by calculating the stratification at the level of the equation of state for a three-parameter, tabulated equation of state.

This paper is organised as follows. In Sec.~\ref{sec:Thermo}, we lay out the thermodynamical framework that forms the basis of three-parameter nuclear-matter equations of state---connecting realistic equations of state with the barotropic models that are usually implemented. The mode calculation is detailed in Sec.~\ref{sec:Pulsation}. In this section, we explicitly show how the matter enters the mathematical formulation of the problem. Adopting a finite-temperature nuclear-matter model in Sec.~\ref{sec:Implement}, we compute the relevant thermodynamics directly from the equation of state and present oscillation results for a low-temperature neutron star. We also consider two thermal models, including a neutron star with uniform redshifted temperature and one with constant entropy per baryon, and compare the two cases. Finally, we summarise and conclude in Sec.~\ref{sec:Conc}.

In this article, we use a positive metric signature $(-, +, +, +)$ and adopt the Einstein summation convention, where repeated indices denote a summation. We use early Latin characters $a, b, c, \ldots$ for spacetime indices and later symbols $j, k, l \ldots$ to represent spatial indices. The labels $(\ell, \m)$ represent the spherical-harmonic numbers. Unless specified otherwise, we will assume natural units, where $G = c = k_\text{B} = 1$.

\section{\label{sec:Thermo}Thermodynamics and the equation of state}

We begin with a discussion on the thermodynamics of the nuclear matter (for additional comprehensive discussions on this topic, see Refs.~\cite{2015PPN....46..633T,2017RvMP...89a5007O,2022EPJA...58..221C}). This is a natural place to start since the thermodynamics encode the microscopic nuclear interactions and determine the material properties of neutron stars. We want to examine the information we need in order to construct realistic, finite-temperature stellar models and how this relates to the fundamental thermodynamics.

Consider a small element of fluid in the neutron star from the point of view of a local inertial reference frame that moves along with the fluid.%
\footnote{The fluid element is small in the sense that its extensive variables are infinitesimal relative to their totals across the neutron star.}
The thermodynamical state of the fluid parcel is fully characterised by its extensive variables---total energy $E$, entropy $S$, proper volume $V$ and particle numbers $\{ N_\text{x} \}$, where $\text{x}$ labels the particle species---and intensive variables---temperature $T$, pressure $p$ and particle chemical potentials $\{ \mu_\text{x} \}$. An infinitesimal change in the total energy $dE$ of the fluid element is related to variations in the other extensive variables $(dS, dV, \{ dN_\text{x} \})$ through the first law of thermodynamics. In the inertial frame, the first law can be expressed as \cite{2021LRR....24....3A}%
\begin{equation}
    dE = T \, dS - p \, dV + \sum_\text{x} \mu_\text{x} \, dN_\text{x}.
\end{equation}
By virtue of the fluid forming an extensive system, we also have the corresponding Euler relation
\begin{equation}
    E = T S - p V + \sum_\text{x} \mu_\text{x} N_\text{x}.
\end{equation}

In order to make progress in describing a neutron star, we must depart from generality and make assumptions about the nuclear composition. For our purposes, we will assume the fluid comprises neutrons, protons and electrons, with respective particle numbers $\{ N_\text{n}, N_\text{p}, N_\text{e} \}$ and chemical potentials $\{ \mu_\text{n}, \mu_\text{p}, \mu_\text{e} \}$ in the fluid element.%
\footnote{Although muons are expected to be present in realistic neutron-star matter, we neglect them here since the nuclear-matter model we adopt later on only considers pure neutron-proton-electron matter. Including muons in the thermodynamics is, however, not difficult since they enter the problem in the same manner as the electrons.}
We also assume that the fluid parcel is charge neutral $N_\text{p} = N_\text{e}$ and contains the number of baryons $N_\text{b} = N_\text{n} + N_\text{p}$. In the context of fluid dynamics, it is convenient to express the thermodynamics in terms of densities.%
\footnote{In doing so, we also reduce the number of independent variables by one. This is a generic feature of an extensive system.}
Therefore, we write
\begin{gather}
    d\varepsilon = n_\text{b} T \, ds + \frac{\varepsilon + p}{n_\text{b}} \, dn_\text{b} + 
    n_\text{b} \mu_\Delta \, dY_\text{e},
    \label{eq:FirstLaw} \\
    \varepsilon = n_\text{b} T s - p + n_\text{b} (\mu_\text{n} + \mu_\Delta Y_\text{e}),
    \label{eq:EulerRelation}
\end{gather}
where $\varepsilon = E / V$ is the total energy density, $s = S / N_\text{b}$ is the entropy per baryon, $n_\text{b} = N_\text{b} / V$ is the number density of baryons, $Y_\text{e} = N_\text{e} / N_\text{b}$ is the electron fraction and $\mu_\Delta = \mu_\text{p} + \mu_\text{e} - \mu_\text{n}$ quantifies deviations from cold $\beta$-equilibrium. Here, we have ignored the effect of neutrinos, since equation-of-state tables assume that they are not trapped in the matter. Effectively, their chemical potentials vanish. For more realistic, hot stars (such as in mergers and proto-neutron stars), neutrinos cannot be assumed to be in thermal equilibrium, and the condition on weak chemical equilibrium is more involved \cite{2018PhRvC..98f5806A}, but we ignore this issue here.

The first law~\eqref{eq:FirstLaw} shows that the system possesses a fundamental potential $\varepsilon = \varepsilon(s, n_\text{b}, Y_\text{e})$ from which all the thermodynamical information can be derived. This is illustrated in the following. From Eq.~\eqref{eq:FirstLaw}, we see that the potentials $T$, $p$ and $\mu_\Delta$ can be directly obtained as partial derivatives of $\varepsilon$. These are
\begin{equation}
    T(s, n_\text{b}, Y_\text{e}) = \frac{1}{n_\text{b}} \left( \frac{\partial \varepsilon}{\partial s} \right)_{n_\text{b}, Y_\text{e}}, \qquad p(s, n_\text{b}, Y_\text{e}) = n_\text{b}^2 \left[ \frac{\partial (\varepsilon / n_\text{b})}{\partial n_\text{b}} \right]_{s, Y_\text{e}}, \qquad \mu_\Delta(s, n_\text{b}, Y_\text{e}) = \frac{1}{n_\text{b}} \left( \frac{\partial \varepsilon}{\partial Y_\text{e}} \right)_{s, n_\text{b}}.
\end{equation}
Then, the final state variable $\mu_\text{n}$ is found using the Euler relation~\eqref{eq:EulerRelation}, which can be expressed as
\begin{equation}
    \mu_\text{n}(s, n_\text{b}, Y_\text{e}) = - \frac{s}{n_\text{b}} \left( \frac{\partial \varepsilon}{\partial s} \right)_{n_\text{b}, Y_\text{e}} + \left( \frac{\partial \varepsilon}{\partial n_\text{b}} \right)_{s, Y_\text{e}} - \frac{Y_\text{e}}{n_\text{b}} \left( \frac{\partial \varepsilon}{\partial Y_\text{e}} \right)_{s, n_\text{b}}.
\end{equation}
This serves as a demonstration of how all the thermodynamical information sits in $\varepsilon = \varepsilon(s, n_\text{b}, Y_\text{e})$. For this reason, $\varepsilon = \varepsilon(s, n_\text{b}, Y_\text{e})$ is the \textit{equation of state} of the nuclear matter that connects the microphysics to the local fluid behaviour. We note that there are other, physically equivalent forms of the equation of state, where different thermodynamic quantities are chosen for the dependent and independent variables. Indeed, later on, we will adopt the free-energy density $f = f(T, n_\text{b}, Y_\text{e})$ (frequently used for finite-temperature nuclear matter) as the fundamental thermodynamic potential and in Footnote~\ref{foot:Enthalpy} below we consider the enthalpy per baryon $h = h(s, p, Y_\text{e})$; both are equivalent forms of the equation of state.

In many neutron-star calculations, further assumptions about the nuclear physics are made. For mature, isolated neutron stars, it is common to treat the matter as cold, $T = 0$, which reduces the number of independent variables by one and leads to a two-parameter equation of state $\varepsilon = \varepsilon(n_\text{b}, Y_\text{e})$. A second common assumption is that the matter maintains (cold) $\beta$ equilibrium, $\mu_\Delta = 0$. This equilibrium corresponds to a balance in the weak nuclear interactions, including electron capture and neutron decay \cite{2018PhRvC..98f5806A}. Chemical equilibrium is a reasonable approximation when the dynamics are much slower than the nuclear-reaction timescales, but will inevitably break down in the opposite regime \cite{1992ApJ...395..240R,2019MNRAS.489.4043A}. Under these assumptions, we end up with
\begin{equation}
    d\varepsilon = \frac{\varepsilon + p}{n_\text{b}} \, dn_\text{b} = \mu_\text{n} \, dn_\text{b}
    \label{eq:Barotropic}
\end{equation}
and thus the fluid is \textit{barotropic}---it depends on only one variable, $\varepsilon = \varepsilon(n_\text{b})$. Under these assumptions, we can freely invert $p = p(n_\text{b})$ to write the energy density as $\varepsilon = \varepsilon[n_\text{b}(p)] = \varepsilon(p)$.%
\footnote{\label{foot:Enthalpy}%
Another way to retrieve this function without an inversion is to make the pressure $p$ a natural variable of the system. This can be achieved with a Legendre transform by introducing the enthalpy per baryon $h = (\varepsilon + p) / n_\text{b}$. Under this transformation, the equation of state is $h = h(s, p, Y_\text{e})$, where $p$ has replaced the baryon-number density $n_\text{b}$ as an independent variable. Thence, the energy density follows from $\varepsilon(s, p, Y_\text{e}) = - p + n_\text{b}(s, p, Y_\text{e}) h(s, p, Y_\text{e})$, which reduces to $\varepsilon = \varepsilon(p)$ for cold, $\beta$-equilibrated matter.}
The latter function is commonly encountered in the astrophysical literature since it closes the structure equations [see Eqs.~\eqref{eqs:Structure} below] and is used (for example) to determine the (static) tidal Love numbers \cite{2008ApJ...677.1216H,2009PhRvD..80h4018B,2009PhRvD..80h4035D}. Indeed, recent constraints on the function $\varepsilon = \varepsilon(p)$ have been obtained from the inspiral of GW170817 \cite{2017PhRvL.119p1101A,2018PhRvL.121p1101A,2019PhRvX...9a1001A}, along with complementary information gleaned from X-ray data \cite{2019ApJ...887L..22R,2020ApJ...893L..21R,2021ApJ...918L..29R}.

While this simplified view of the nuclear matter is perfectly appropriate for describing cold, equilibrium neutron stars, it is important to note that a relation $\varepsilon = \varepsilon(p)$ is in fact insufficient to describe the \textit{full} thermodynamical state of the star. (We illustrate and elaborate on this point later.) Our objective should be to obtain the fundamental thermodynamical description at the level of the equation of state.

More severely, the assumptions of $\mu_\Delta = T = 0$ will fail for some physical situations of interest---three of which immediately come to mind: pulsating neutron stars (the focus of this article), binary mergers (which we explore in Ref.~\cite{2025PhRvD.111b3049G}) and proto-neutron stars. In all these cases, the dynamical timescales may be sufficiently fast such that $\beta$ equilibrium can no longer be assumed \cite{2021PhRvD.104j3006H,2023PhRvD.107d3023H}. Furthermore, the mergers and supernovae that we wish to examine with gravitational-wave and electromagnetic instruments are hot, energetic events. For these reasons, realistic calculations of neutron-star dynamics must go beyond these assumptions.

\section{\label{sec:Pulsation}The pulsation problem}

\subsection{The equilibrium neutron star}

With the thermodynamics accounted for, we are now in a position to connect the equation of state with the stellar model. We treat the neutron star as a non-rotating, perfect fluid. The background is spherically symmetric and its geometry is characterised by the metric $g_{a b}$, given by the usual line element
\begin{equation}
    ds^2 = g_{a b} \, dx^a dx^b = - e^\nu \, dt^2 + e^\lambda \, dr^2 + r^2 \, (d\theta^2 + \sin^2 \theta \, d\phi^2),
\end{equation}
with time coordinate $t$, spherical polar coordinates $(r, \theta, \phi)$ and metric potentials $\nu(r)$ and $\lambda(r)$. By continuity with the exterior Schwarzschild spacetime, the gravitational mass $m(r)$ enclosed in areal radius $r$ is defined by
\begin{equation}
    e^\lambda = \frac{1}{1 - 2 m / r}.
\end{equation}
In the proper reference frame, the perfect fluid is described by the conservative stress-energy tensor
\begin{equation}
    T_{a b} = (\varepsilon + p) u_a u_b + p g_{a b},
\end{equation}
where the fluid four-velocity $u^a$ has components
\begin{equation}
    u^t = e^{- \nu / 2}, \qquad u^j = 0.
\end{equation}
Finally, the Einstein field equations lead to the well-known relativistic equations of stellar structure
\begin{equation}
    \frac{dm}{dr} = 4 \pi r^2 \varepsilon, \qquad \frac{d\nu}{dr} = 2 \frac{m + 4 \pi r^3 p}{r (r - 2 m)}, \qquad \frac{dp}{dr} = - \frac{\varepsilon + p}{2} \frac{d\nu}{dr}.
    \label{eqs:Structure}
\end{equation}
Equations~\eqref{eqs:Structure} can be integrated provided a relationship between the energy density $\varepsilon$ and pressure $p$. From the solution to Eqs.~\eqref{eqs:Structure}, one obtains the stellar radius $R$ and total gravitational mass $M = m(R)$, which characterise the exterior spacetime of the spherical star.

At this point, we pick up on the issue of the equation of state (as introduced in Sec.~\ref{sec:Thermo}) and how it enters the macroscopic description of the neutron star. For this discussion, we will assume cold, $\beta$-equilibrated nuclear matter. Recall that, in this regime, the fluid is barotropic and the thermodynamical system can be fully characterised by the equation of state $\varepsilon = \varepsilon(n_\text{b})$ [see Eq.~\eqref{eq:Barotropic}]. From the equation of state, we can straightforwardly derive the thermodynamical potentials $p = p(n_\text{b})$ and $\mu_\text{n} = \mu_\text{n}(n_\text{b})$.

Although $\varepsilon = \varepsilon(p)$ is the only relation required from the thermodynamics to solve Eqs.~\eqref{eqs:Structure}, it is not sufficient to describe the complete thermodynamical state of the neutron star. For example, suppose we wanted to know the distribution of baryons $N_\text{b}(r)$ in the star. This would be necessary to find the total number of baryons $N_\text{b}(R)$ and the star's baryonic mass $M_\text{b} = m_\text{b} N_\text{b}(R)$, where $m_\text{b}$ is the mass of a baryon. To determine $N_\text{b}(r)$, we must integrate the corresponding number density $n_\text{b}$ over the proper volume of the star. This is equivalent to solving the differential equation
\begin{equation}
    \frac{dN_\text{b}}{dr} = \frac{4 \pi r^2 n_\text{b}}{\sqrt{1 - 2 m / r}}.
    \label{eq:BaryonNumber}
\end{equation}
Here the issue is apparent: to solve Eq.~\eqref{eq:BaryonNumber}, we require a thermodynamical relation in addition to $\varepsilon = \varepsilon(p)$. A function $n_\text{b} = n_\text{b}(p)$ [or equivalently $n_\text{b} = n_\text{b}(\varepsilon)$] is necessary to close Eq.~\eqref{eq:BaryonNumber}. By thermodynamical consistency, one can integrate Eq.~\eqref{eq:Barotropic} to obtain $n_\text{b} = n_\text{b}(p)$, but only up to an integration constant. The resolution to this is that both functions would follow naturally from a fundamental thermodynamical potential, such as $\varepsilon = \varepsilon(n_\text{b})$. Thus, while $\varepsilon = \varepsilon(p)$ is perfectly sufficient to solve Eqs.~\eqref{eqs:Structure}, it cannot describe the full thermodynamics of the star.

Before we move on, it is worth briefly commenting on the physics that cannot be extracted in the barotropic regime, even with the fundamental potential $\varepsilon = \varepsilon(n_\text{b})$. To illustrate this, we relax the previous assumptions and assume a three-dimensional equation of state, where we can freely compute the state variables. From this starting point, if we now choose to consider zero-temperature, equilibrium matter, we may solve the coupled, non-linear system of equations $T(s, n_\text{b}, Y_\text{e}) = 0$ and $\mu_\Delta(s, n_\text{b}, Y_\text{e}) = 0$ to obtain $s = s(n_\text{b})$ and $Y_\text{e} = Y_\text{e}(n_\text{b})$. However, under the barotropic assumption, we observe that the first law~\eqref{eq:Barotropic} no longer contains information on the entropy and chemical abundances in the fluid. Hence, this information will not be accessible to us if we only focus our attention on constraining $\varepsilon = \varepsilon(n_\text{b})$. (We note that there have been some recent discussion on this issue in the context of gravitational-wave astronomy \cite{2020ApJ...899....4X,2022PhRvD.105h3016M,2023PhRvD.108l2006I,2024PhRvC.109b5804I}.) Fortunately, dynamical contexts involve more physics and probe a different regime.

\subsection{The oscillations}

Let us move on to the natural vibrations of the neutron star. To calculate the oscillation modes, we will use the formulation of the perturbation equations of Detweiler and Lindblom~\cite{1983ApJS...53...73L,1985ApJ...292...12D}, along with the augmentation introduced by \citet{2015PhRvD..92f3009K} to address numerical noise in the low-frequency spectrum. For the exterior problem, we use the method of \citet{1995MNRAS.274.1039A}. (We refer the interested reader to these papers for the full system of equations.) Much of this work rests on the framework of relativistic Lagrangian perturbation theory for fluids \cite{1978CMaPh..62..247F,2013rrs..book.....F,2021LRR....24....3A}. The numerical implementation that supports the results in this paper is publicly available \cite{code} and has been used to support a recent study \cite{2025PhRvD.111b3049G}. To verify the robustness of our code, we compared it to results in Refs.~\cite{1995MNRAS.274.1039A,1992MNRAS.255..119K,2015PhDT.......604K} and found good agreement.

The linear perturbations of a relativistic star can be described by the Eulerian perturbation of the metric $h_{a b}$ and the Lagrangian displacement vector of the fluid $\xi^a$. Since our focus is on the non-radial oscillations of a spherically symmetric neutron star, we can decompose the perturbations into the form $Y_\ell^\m e^{i \omega t}$, where $\omega$ is the natural frequency of a mode and $Y_\ell^\m(\theta, \phi)$ is a spherical harmonic with quantum numbers $(\ell, \m)$. For the spacetime, we adopt the Regge-Wheeler gauge, where high angular derivatives are chosen to vanish \cite{1957PhRv..108.1063R}. In the polar sector,%
\footnote{It is well known that (with the exception of the rapidly damped gravitational-wave \textit{w}-modes \cite{1994MNRAS.268.1015K}) the axial perturbations of a non-rotating, fluid star are trivial solutions to the perturbation problem (see, \textit{e.g.}, Refs.~\cite{1999ApJ...521..764L,2000PhRvD..63b4019L}). Other axial oscillation modes exist in instances where there is anisotropy in the neutron-star structure, such as rotation \cite{2000PhRvD..63b4019L} or the presence of an elastic crust \cite{2015PhRvD..92f3009K}. For this calculation, we can ignore them.}
the metric perturbation reads
\begin{equation}
    [h_{a b}] = -
    \begin{bmatrix}
        e^\nu r^\ell H_0 & i \omega r^{\ell + 1} H_1 & 0 & 0 \\
        i \omega r^{\ell + 1} H_1 & e^\lambda r^\ell H_2 & 0 & 0 \\
        0 & 0 & r^{\ell + 2} K & 0 \\
        0 & 0 & 0 & r^{\ell + 2} \sin^2 \theta K
    \end{bmatrix}
    Y_\ell^\m e^{i \omega t},
\end{equation}
which depends on the functions $H_0(r)$, $H_1(r)$, $H_2(r)$ and $K(r)$. The displacement vector describes how the fluid elements in the equilibrium configuration are displaced in the perturbation. We can exploit the gauge freedom associated with the Lagrangian formulation to set $u_a \xi^a = 0$ \cite{1977AnPhy.107....1S} and decompose the non-trivial components as
\begin{equation}
    \xi^r = e^{- \lambda / 2} r^{\ell - 1} W Y_\ell^\m e^{i \omega t}, \qquad \xi^\theta = - r^{\ell - 2} V \partial_\theta Y_\ell^\m e^{i \omega t}, \qquad \xi^\phi = - \frac{r^{\ell - 2}}{\sin^2 \theta} V \partial_\phi Y_\ell^\m e^{i \omega t},
\end{equation}
with functions $W(r)$ and $V(r)$. The perturbation is described by the spacetime $(H_0, H_1, H_2, K)$ and fluid functions $(W, V)$. However, some additional simplifications can be made.

\citet{1985ApJ...292...12D} reduced the pulsation equations (initially derived by \citet{1967ApJ...149..591T}) for the interior of the star to a system of four first-order differential equations for the functions $(H_1, K, W, X)$, where $X$ is an auxiliary variable proportional to the Lagrangian pressure perturbation. These equations correspond to the two degrees of freedom in the fluid and two for the gravitational radiation and are solved subject to boundary conditions. Only two of the four linearly independent solutions satisfy regularity at the stellar centre. At the surface, a physically acceptable solution must have a vanishing Lagrangian perturbation of the pressure, $X(R) = 0$, which three linearly independent solutions can satisfy. These boundary conditions are sufficient to determine $(H_1, K, W, X)$ in the stellar interior for any given frequency $\omega$ (up to an arbitrary amplitude).

A generic perturbation of the star at frequency $\omega$ will have ingoing and outgoing gravitational waves at infinity. Indeed, ingoing radiation with frequency  $\omega$ will cause the neutron star to oscillate at that frequency. Therefore, a free oscillation mode of the star corresponds to a solution with purely outgoing waves.

In the exterior, the functions associated with the fluid vanish and we only have perturbations of the spacetime. The metric perturbation functions $H_1$ and $K$ can be combined into a single differential equation, known as the \textit{Zerilli equation} \cite{1970PhRvD...2.2141Z,1971ApJ...166..197F}. This second-order equation represents the two degrees of freedom in the gravitational field---ingoing and outgoing radiation. An oscillation mode corresponds to when the amplitude of ingoing gravitational waves vanishes. This defines an eigenvalue problem, with eigenfrequency $\omega$.

Just as the background Einstein equations needed a relation connecting $\varepsilon$ and $p$, the perturbations require a relationship between the (Lagrangian) perturbations $\Delta \varepsilon$ and $\Delta p$. This information comes from the thermodynamics of the nuclear matter. If we continued our assumptions that the matter is cold and the nuclear reactions are sufficiently fast to maintain $\beta$ equilibrium as the star oscillates, then we can assert $\Delta p / p = \Gamma \Delta \varepsilon / (\varepsilon + p)$, where
\begin{equation}
    \Gamma = \frac{\varepsilon + p}{p} \frac{d p}{d \varepsilon}.
    \label{eq:Gammaa}
\end{equation}
This is the barotropic assumption, commonly adopted in the literature and used in the original neutron-star mode calculations of Detweiler and Lindblom~\cite{1983ApJS...53...73L,1985ApJ...292...12D}. The index $\Gamma$ can be immediately inferred from the function $\varepsilon = \varepsilon(p)$. However, as we have already alluded, this is not the correct regime for mature neutron stars, for which the timescale required to reach chemical equilibrium is slow compared to the fluid oscillations.

It was first argued by \citet{1992ApJ...395..240R} that the characteristic timescale associated with the relevant weak interactions is actually much longer than a typical oscillation period of a neutron star. This means that, as a fluid element in the star is displaced from its equilibrium position during a perturbation, the reactions do not act fast enough to restore $\beta$ equilibrium between the element and its new environment. Thus, bouyancy forces associated with the differing composition make the fluid parcel oscillate. These oscillations are (low-frequency) \textit{g}-modes, sourced by composition gradients.%
\footnote{The compositional \textit{g}-modes join the spectrum alongside the well-known class of \textit{g}-modes that oscillate due to entropy gradients in the stellar material. For a finite-temperature neutron star, we expect both effects to contribute to the buoyancy for \textit{g}-modes to exist.}
Therefore, for neutron stars, it is instead more appropriate to regard the chemical composition of the fluid as \textit{frozen} during an oscillation, $\Delta Y_\text{e} = 0$. (See Ref.~\cite{2024MNRAS.531.1721C} for a recent study on the \textit{g}-mode spectrum in the intermediate regime of finite reaction times.)

Assuming that no heat is transferred, $\Delta s = 0$, we have
\begin{equation}
    \frac{\Delta p}{p} = \Gamma_1 \frac{\Delta n_\text{b}}{n_\text{b}} = \Gamma_1 \frac{\Delta \varepsilon}{\varepsilon + p},
    \label{eq:ThermoPert}
\end{equation}
where
\begin{equation}
    \Gamma_1 = \left( \frac{\partial \ln p}{\partial \ln n_\text{b}} \right)_{s, Y_\text{e}} = \frac{\varepsilon + p}{p} \left( \frac{\partial p}{\partial \varepsilon} \right)_{s, Y_\text{e}}
    \label{eq:Gamma1a}
\end{equation}
is the adiabatic index, which holds information about composition and entropy gradients in the nuclear matter. Motivated by Eq.~\eqref{eq:ThermoPert}, it is useful to introduce the (relativistic) Schwarzschild discriminant
\begin{equation}
    A = \frac{1}{\varepsilon + p} \frac{d\varepsilon}{dr} - \frac{1}{\Gamma_1 p} \frac{dp}{dr} = \left( \frac{1}{\Gamma} - \frac{1}{\Gamma_1} \right) \frac{d\ln p}{dr},
\end{equation}
which determines the convective stability of the star \cite{1973ApJ...185..685D}. If $A \leq 0$ everywhere in the fluid (equivalently, $\Gamma_1 \geq \Gamma$), then the star is convectively stable and supports \textit{g}-mode oscillations. Otherwise, if $A = 0$ (when $\Gamma_1 = \Gamma$), the \textit{g}-modes instead reside in the zero-frequency subspace of perturbations \cite{1999ApJ...521..764L,2000PhRvD..63b4019L}.

\section{\label{sec:Implement}Implementing a realistic matter model}

To illustrate the steps involved in incorporating a realistic description of the nuclear matter, we adopt the finite-temperature APR model \cite{2019PhRvC.100b5803S} as our nuclear-matter equation of state (as implemented in the {\footnotesize CompOSE} library \cite{2015PPN....46..633T,2017RvMP...89a5007O,2022EPJA...58..221C}). The current state of the art is based on three-parameter models with temperature $T$, baryon-number density $n_\text{b}$ and fraction of electrons $Y_\text{e}$ as the natural variables of the system \cite{2017RvMP...89a5007O}.%
\footnote{The equations of state in {\scriptsize CompOSE} in general use the charge fraction of strongly interacting particles $Y_\text{q}$ as an independent variable. This is because this quantity is well-defined for pure hadronic matter and extends naturally to incorporate muons. For the composition we assume, $Y_\text{q} = Y_\text{e}$.}
This is accomplished by using the (Helmholtz) free-energy density $f = \varepsilon - n_\text{b} T s$ as the fundamental potential. With this potential, the first law~\eqref{eq:FirstLaw} and Euler relation~\eqref{eq:EulerRelation} become
\begin{gather}
    df = - n_\text{b} s \, dT + \frac{f + p}{n_\text{b}} \, dn_\text{b} + n_\text{b} \mu_\Delta \, dY_\text{e}, \\
    f = - p + n_\text{b} (\mu_\text{n} + \mu_\Delta Y_\text{e})
\end{gather}
and $f = f(T, n_\text{b}, Y_\text{e})$ is the equation of state. In practice, the relevant quantities that describe the matter's thermodynamical state are usually provided in tabulated form. We now describe how one calculates a pulsating star from a three-parameter equation-of-state table.

We start from the table, which includes the pressure $p$, entropy per baryon $s$, deviation from $\beta$ equilibrium $\mu_\Delta$ and total energy density $\varepsilon$ on a grid of $(T, n_\text{b}, Y_\text{e})$. We interpolate this information to find the functions $p = p(T, n_\text{b}, Y_\text{e})$, $s = s(T, n_\text{b}, Y_\text{e})$, $\mu_\Delta = \mu_\Delta(T, n_\text{b}, Y_\text{e})$ and $\varepsilon = \varepsilon(T, n_\text{b}, Y_\text{e})$.%
\footnote{To account for the wide spread in magnitudes, we actually use logarithms to interpolate the quantities $\ln p$, $\ln s$, $\mu_\Delta$ and $\ln \varepsilon$ as functions of $(\ln T, \ln n_\text{b}, Y_\text{e})$ using splines. For the APR model that we work with, the grid is given as $0.00999 \leq T / \unit{\mega\electronvolt} \leq 251$, $\num{5.43e-13} \leq n_\text{b} / \unit{\per\femto\metre\cubed} \leq 6.32$ and $0.005 \leq Y_\text{e} \leq 0.655$.}
In addition to the functions, we also require their partial derivatives to determine the indices $\Gamma = \Gamma(T, n_\text{b}, Y_\text{e})$ and $\Gamma_1 = \Gamma_1(T, n_\text{b}, Y_\text{e})$. For this reason, we use splines to carry out the interpolation, where it is trivial to obtain the partial derivatives. We use Eq.~\eqref{eq:Gammaa} to write
\begin{equation}
    \Gamma = \frac{\varepsilon + p}{p} \frac{dp / dn_\text{b}}{d\varepsilon / dn_\text{b}},
\end{equation}
where the derivatives are given by
\begin{align}
    \frac{dp}{dn_\text{b}} &= \left( \frac{\partial p}{\partial T} \right)_{n_\text{b}, Y_\text{e}} \frac{dT}{dn_\text{b}} + \left( \frac{\partial p}{\partial n_\text{b}} \right)_{T, Y_\text{e}} + \left( \frac{\partial p}{\partial Y_\text{e}} \right)_{T, n_\text{b}} \frac{dY_\text{e}}{dn_\text{b}}, \label{eq:dp_dnb} \\
    \frac{d\varepsilon}{dn_\text{b}} &= \left( \frac{\partial \varepsilon}{\partial T} \right)_{n_\text{b}, Y_\text{e}} \frac{dT}{dn_\text{b}} + \left( \frac{\partial \varepsilon}{\partial n_\text{b}} \right)_{T, Y_\text{e}} + \left( \frac{\partial \varepsilon}{\partial Y_\text{e}} \right)_{T, n_\text{b}} \frac{dY_\text{e}}{dn_\text{b}}.
\end{align}
For Eq.~\eqref{eq:Gamma1a}, we use $\Delta s = \Delta Y_\text{e} = 0$ to infer
\begin{equation}
    \Gamma_1 = \frac{n_\text{b}}{p} \left[ \left( \frac{\partial p}{\partial n_\text{b}} \right)_{T, Y_\text{e}} - \left( \frac{\partial p}{\partial T} \right)_{n_\text{b}, Y_\text{e}} \left( \frac{\partial s}{\partial n_\text{b}} \right)_{T, Y_\text{e}} \left( \frac{\partial s}{\partial T} \right)_{n_\text{b}, Y_\text{e}}^{-1} \right].
    \label{eq:Gamma1b}
\end{equation}
We obtain the partial thermodynamic derivatives in $\Gamma$ and $\Gamma_1$ from the interpolation functions.%
\footnote{It is worth noting that second derivatives of the fundamental potential are not independent. They are related via the Maxwell relations. The Maxwell relation that simplifies Eq.~\eqref{eq:Gamma1b} is $(\partial p / \partial T)_{n_\text{b}, Y_\text{e}} = - n_\text{b}^2 (\partial s / \partial n_\text{b})_{T, Y_\text{e}}$.}
For this reason, it is important to use an interpolation routine that ensures continuity of the functions and their derivatives in the grid. It should be noted that the partial derivatives are not explicitly provided in the table. Thus, a numerical technique to approximate the derivatives is necessary. This is simple to carry out using splines.

Next, we assume the background star is in chemical equilibrium to determine the composition. This is carried out by solving
\begin{equation}
    \mu_\Delta(T, n_\text{b}, Y_\text{e}) = 0
    \label{eq:BetaEquilibrium}
\end{equation}
to obtain the function $Y_\text{e} = Y_\text{e}(T, n_\text{b})$. Here, we recall that this condition is sufficient for $\beta$ equilibrium only when the star is cold. For appreciable temperatures, the neutrinos are no longer in thermal equilibrium and Eq.~\eqref{eq:BetaEquilibrium} is invalid \cite{2018PhRvC..98f5806A}. To address this issue in practice, the neutrino dynamics must be incorporated in a separate (and often computationally demanding) fashion. As already mentioned, we ignore this issue here. Although Eq.~\eqref{eq:BetaEquilibrium} represents a simple root-search, it must be carried out at every integration step in the background and perturbation equations. Thus, to speed up the computations, we determine $Y_\text{e}$ at each $(T, n_\text{b})$ in the original grid and pass it to the interpolation functions---\textit{e.g.}, $p = p[T, n_\text{b}, Y_\text{e}(T, n_\text{b})] = p(T, n_\text{b})$---to form a two-dimensional table for the matter model. (A similar approach is  adopted in some numerical-relativity simulations of neutron-star mergers.) For the APR model, we find that, for certain values of $(T, n_\text{b})$, $\mu_\Delta$ does not change sign with varying $Y_\text{e}$. We therefore restricted the two-dimensional grid to $0.01 \lesssim T / \unit{\mega\electronvolt} \lesssim 230$ and $\num{1.3e-7} \lesssim n_\text{b} / \unit{\per\femto\metre\cubed} \lesssim 2.5$, where equilibrium can be enforced.

Then we interpolate the quantities on the new, two-dimensional grid. To calculate the equilibrium star, we invert $p = p(T, n_\text{b})$ to obtain $n_\text{b} = n_\text{b}(T, p)$. Given the temperature $T$, this is sufficient information to integrate the structure equations~\eqref{eqs:Structure}, as well as Eq.~\eqref{eq:BaryonNumber}. The perturbation equations need $\Gamma = \Gamma(T, n_\text{b})$ and $\Gamma_1 = \Gamma_1(T, n_\text{b})$. To determine $\Gamma$, we require
\begin{equation}
    \frac{dY_\text{e}}{dn_\text{b}} = \left( \frac{\partial Y_\text{e}}{\partial T} \right)_{n_\text{b}} \frac{dT}{dn_\text{b}} + \left( \frac{\partial Y_\text{e}}{\partial n_\text{b}} \right)_{T},
    \label{eq:dYe_dnb}
\end{equation}
which is inferred from $Y_\text{e} = Y_\text{e}(T, n_\text{b})$. This concludes the description of how a three-parameter matter model enters the perturbation problem.

\subsection{A cold neutron star}

\begin{figure}
    \includegraphics[width=0.5\textwidth]{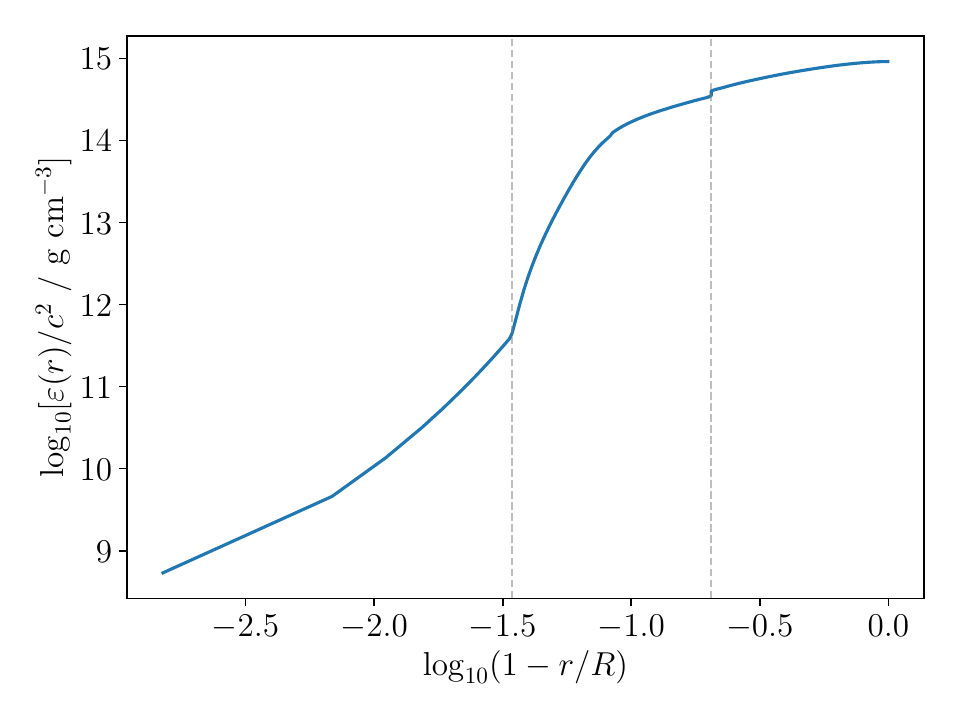}%
    \includegraphics[width=0.5\textwidth]{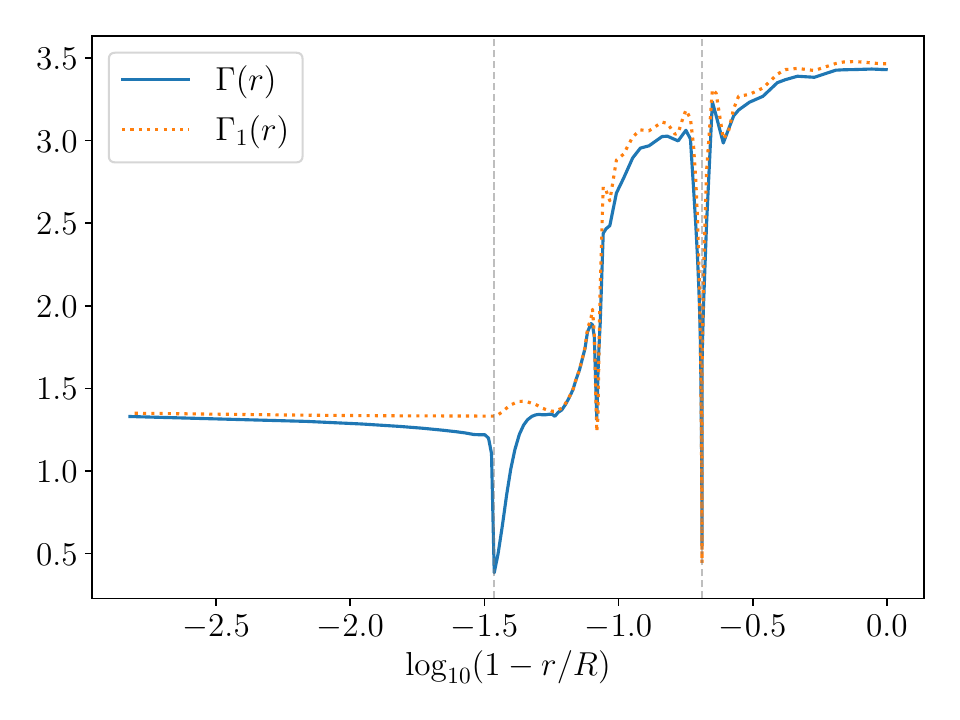}
    \caption{\label{fig:Profiles}%
    The profiles of the energy density (left panel) and adiabatic indices (right panel) in the cold neutron star described by the APR equation of state at fixed $T = \qty{0.02}{\mega\electronvolt}$. The two vertical, dashed lines in each panel correspond to the transition between nucleons to a pion condensate in the APR model at high density (right) and neutron drip at low density (left). The phase transition produces a discontinuity in the density, which results in sharp behaviour in the indices. Neutron drip sources a change in the density derivative, corresponding to the second sharp feature (at lower densities) in the background index $\Gamma$. The difference between the adiabatic indices quantifies the degree of stratification (composition and entropy gradients, in general) that support \textit{g}-mode oscillations. At high densities, the composition gradients are relatively weak and so the adiabatic indices are roughly similar. However, at densities surrounding neutron drip a cavity develops, which provides an enhanced buoyancy force, giving rise to the higher frequency \textit{g}\textsubscript{2}-mode.}
\end{figure}

We are now in a position to determine the oscillation modes of a neutron star described by the three-parameter APR model \cite{2019PhRvC.100b5803S}. First, we will explore the $\ell = 2$ oscillation spectrum of a neutron star of $M_\text{b} = 1.4 M_\odot$ with constant $T = \qty{0.02}{\mega\electronvolt}$. Since the temperature is uniform, we have $dT / dn_\text{b} = 0$. Due to the low-temperature cut-off in the table (which is a typical feature in nuclear-matter models), we must assume temperatures above that of mature neutron stars (below $T \sim \qty{0.1}{\kilo\electronvolt} \sim \qty{e6}{\kelvin}$). An obvious alternative to this approach would be to use a zero-temperature matter model with composition. However, since our interest is in laying the foundations for exploring thermal effects (which we do in Sec.~\ref{sec:Temp} and for merger simulations in Ref.~\cite{2025PhRvD.111b3049G}), we choose to work with the low-temperature limit of a finite-temperature equation of state. Still, at the chosen temperature the thermal pressure is fairly weak compared to the fluid contribution and we expect the results to be representative of a thermally cold neutron star. We show in Fig.~\ref{fig:Profiles} the energy density and adiabatic indices of the equilibrium neutron star, indicating the phase transition in the APR model from nucleonic matter (at low densities) to a neutral pion condensate (at high densities), as well as the location of neutron drip. The phase transition occurs at $n_\text{b} \approx \qty{0.2}{\per\femto\metre\cubed}$.

Our calculation is carried out in full general relativity and therefore gravitational radiation is emitted by the modes. The waves carry away energy and damp the oscillations, which result in the eigensolutions being complex. Although we allow for complex numbers in our numerical determination of the perturbations, we will only show the real parts in our results, since the \textit{g}-modes we focus on are weakly damped and thus have small imaginary contributions to the frequency.

\begin{figure}
    \includegraphics[width=0.5\textwidth]{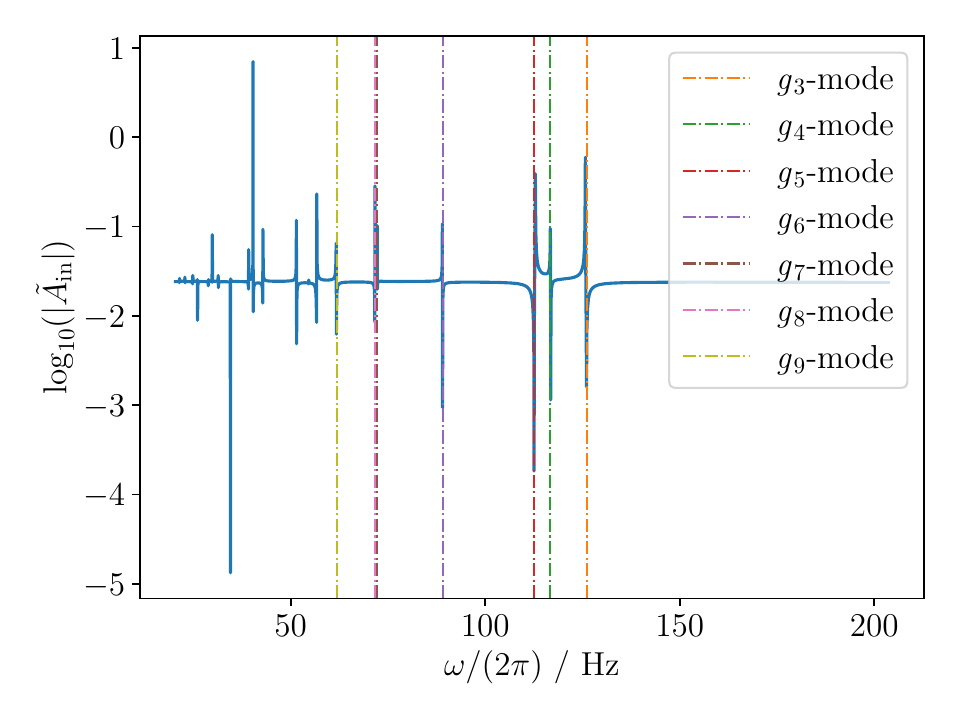}%
    \includegraphics[width=0.5\textwidth]{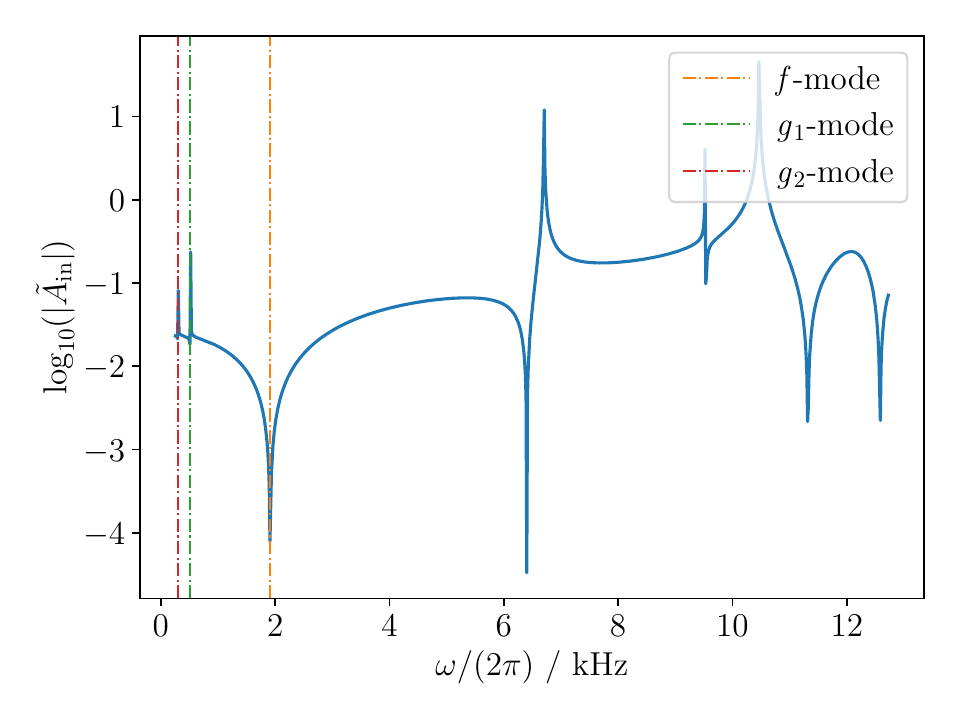}
    \caption{\label{fig:Spectrum}%
    The oscillation spectrum for quadrupolar perturbations of the cold neutron star, visualised with the ingoing gravitational-wave amplitude against (real) oscillation frequency. The left panel focuses on a low-frequency portion of the spectrum, while the right panel displays high frequencies. A narrow singularity corresponds to a slowly damped mode solution. The low-frequency singularities represent the \textit{g}-modes of the star, restored by stratification in the thermal and compositional gradients, and the phase transition. Vertical, dash-dotted lines indicate the real frequencies of the oscillation modes summarised in Table~\ref{tab:Cold}. The two highest frequency \textit{g}-modes are not visible in the low-frequency range, but can be seen in the right panel. The right panel shows the \textit{f}-mode and the first few \textit{p}-modes at successively higher frequencies.}
\end{figure}

In Fig.~\ref{fig:Spectrum}, we show a visualisation of the oscillation spectrum for quadrupolar perturbations. The figure presents the amplitude of ingoing radiation $\tilde{A}_\text{in}$ (with a particular normalisation to flatten out the spectrum) versus real-valued frequencies \cite{1995MNRAS.274.1039A}. A mode solution corresponds to zeros of $\tilde{A}_\text{in}$ and appear as sharp singularities in Fig.~\ref{fig:Spectrum}.

With the spectrum visualised, we are in a position to precisely determine the eigensolutions of the oscillation modes. We start by searching for the fundamental \textit{f}-mode of the star. The \textit{f}-mode is identified by its broad singularity in the spectrum (see the right panel of Fig.~\ref{fig:Spectrum}). We find it with frequency $\text{Re}(\omega) / (2 \pi) = \qty{1.9123}{\kilo\hertz}$. As is characteristic of \textit{f}-modes, the eigenfunction gradually increases away from the centre and is largest at the stellar surface.

\begin{figure}
    \includegraphics[width=0.5\textwidth]{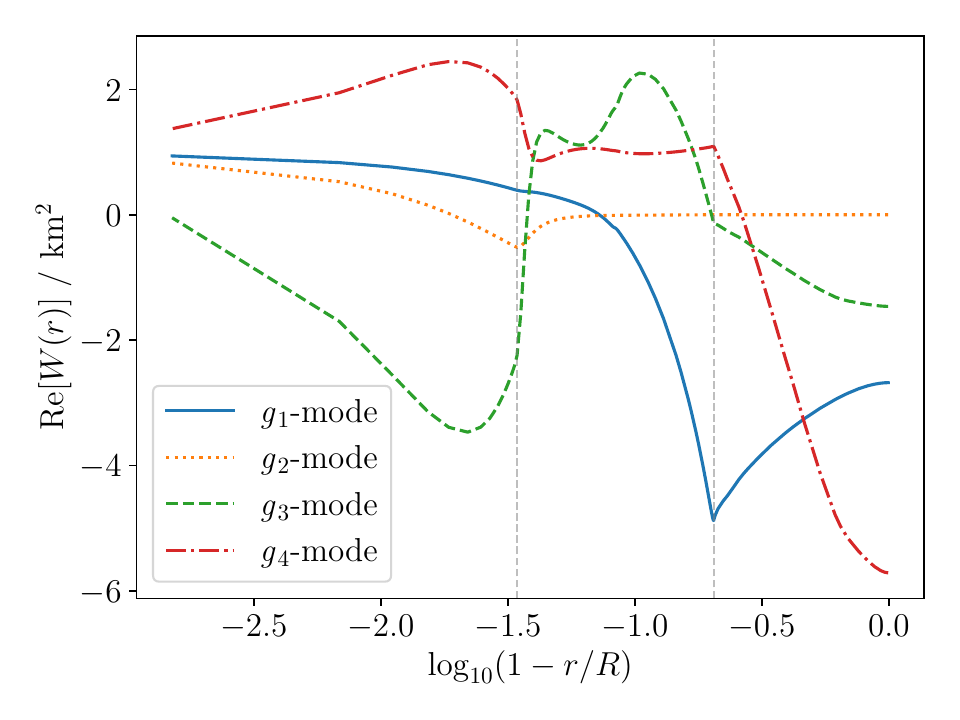}
    \caption{\label{fig:g-modes}%
    The radial-displacement eigenfunction of the first four quadrupolar \textit{g}-modes of the cold neutron star. The vertical, dashed lines correspond to the features indicated in Fig.~\ref{fig:Profiles}. The \textit{g}\textsubscript{1}-mode eigenfunction has a particularly prominent spike at the energy-density discontinuity (the phase transition). Some of the other \textit{g}-modes have notable features at neutron drip.}
\end{figure}

Next, we examine the low-frequency perturbations. We show the eigenfunctions of the first four \textit{g}-modes in Fig.~\ref{fig:g-modes}. The first \textit{g}-mode---the \textit{g}\textsubscript{1}-mode---has frequency $\text{Re}(\omega) / (2 \pi) = \qty{508.5}{\hertz}$. The sharp feature in its eigenfunction indicates its association with the energy-density discontinuity at the phase transition from nucleonic matter to the pion condensate (around $\varepsilon / c^2 \sim \qty{e14}{\gram\per\cubic\centi\metre}$), shown in the left panel of Fig.~\ref{fig:Profiles}. The frequency for this oscillation mode is higher than a typical \textit{g}-mode. We will return to this point in a moment.

\begin{figure}
    \includegraphics[width=0.5\textwidth]{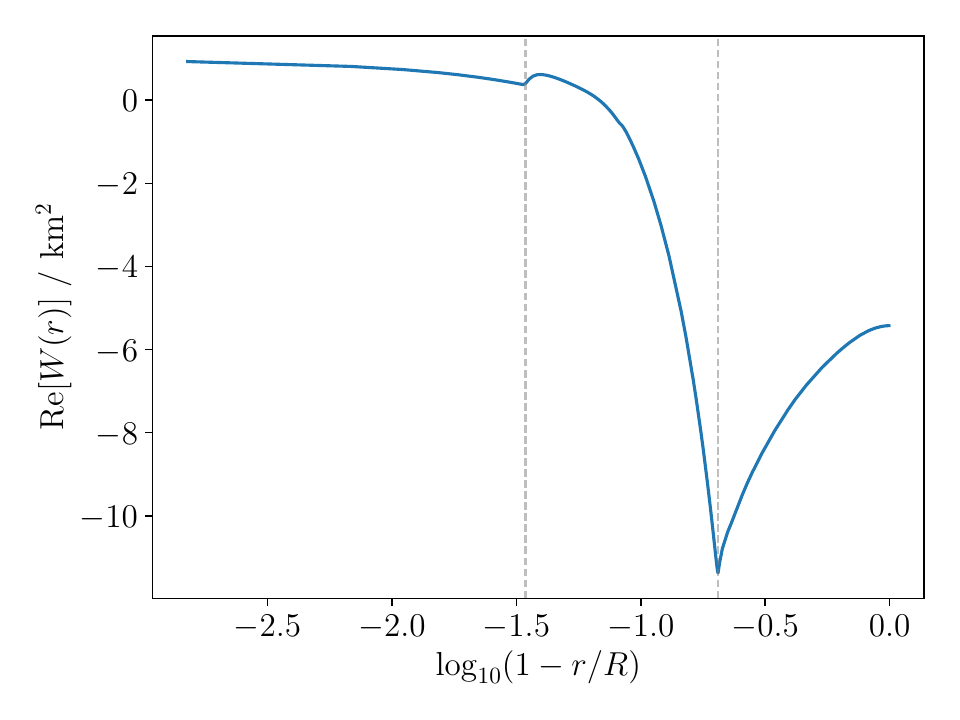}
    \caption{\label{fig:i-mode}%
    The radial-displacement eigenfunction of the quadrupolar \textit{i}-mode of the unstratified, cold neutron star. An \textit{i}-mode is sourced by the presence of a first-order phase transition in the stellar material. For the APR matter model we adopt here, this phase transition occurs at high densities and is a transition from nucleons to a pion condensate. The vertical, dashed lines correspond to the features indicated in Fig.~\ref{fig:Profiles}. We see that the \textit{i}-mode has the characteristic behaviour, where its eigenfunction rises from the stellar centre up to the position of the interface (right vertical, dashed line) and then obtains a kink. The eigenfunction is qualitatively similar to that of the \textit{g}\textsubscript{1}-mode encountered in the realistic, stratified neutron star shown in Fig.~\ref{fig:g-modes}.}
\end{figure}

The \textit{g}-mode perturbations arise when there is stratification in the fluid. As we have discussed, this naturally occurs as the entropy and chemical composition vary in the stellar profile. However, there is another form of stratification that is expected to occur in neutron stars: phase transitions that lead to discontinuities in the energy density. Such a phase transition will exist at the interface between the fluid core and elastic crust, and may arise in the high-density core if there is, for example,  quark deconfinement. These interfaces lead to the presence of \textit{discontinuity} \textit{g}-modes, also commonly known as \textit{i}-modes \cite{1987MNRAS.227..265F,1988ApJ...325..725M,2012PhRvL.108a1102T}. To clarify this identification, we have also computed the mode spectrum when the stratification is artificially removed by setting $\Gamma_1 = \Gamma$. For the unstratified stellar model, the high-frequency portion is largely unchanged; the \textit{f}-mode moves to $\text{Re}(\omega) / (2 \pi) = \qty{1.9122}{\kilo\hertz}$. However, by removing the composition gradients, the entire low-frequency spectrum is eliminated; the only exception to this is a solution with $\text{Re}(\omega) / (2 \pi) = \qty{471.5}{\hertz}$. We see from its eigenfunction, shown in Fig.~\ref{fig:i-mode}, that it has two features that directly correspond to the energy-density profile, including the sharp feature at the phase transition. Thus, the \textit{i}-mode of the unstratified star and the \textit{g}\textsubscript{1}-mode likely have the same origin. However, their eigenfrequencies substantially differ. We will continue referring to the oscillation of the stratified neutron star as the \textit{g}\textsubscript{1}-mode.

\citet{1987MNRAS.227..265F} used a simple, toy model to explore the functional behaviour of the discontinuity modes. In particular, he found that the mode eigenfrequency scales with the density jump and the depth at which the interface occurs. We observe in the left panel of Fig.~\ref{fig:Profiles} that the discontinuity that gives rise to the mode is large. Hence, the large frequency of the \textit{g}\textsubscript{1}-mode can be explained by the relatively large discontinuity in the APR equation of state.

Another (less prominent) feature is seen in the \textit{g}\textsubscript{2}-mode eigenfunction at a different point in the star. This mode has frequency $\text{Re}(\omega) / (2 \pi) = \qty{304.5}{\hertz}$ and the sign change in its eigenfunction derivative corresponds to neutron drip in the crust at $\varepsilon / c^2 \sim \qty{4e11}{\gram\per\cubic\centi\metre}$. At neutron drip, Fig.~\ref{fig:Profiles} shows that the derivative of the energy density is discontinuous, which is seen clearly in the profile of $\Gamma$. The eigenfunctions in Fig.~\ref{fig:g-modes} show that other \textit{g}-modes are also sensitive to this feature. We note that the \textit{g}\textsubscript{2}-mode frequency is somewhat higher than typical \textit{g}-mode oscillations in neutron stars (see, \textit{e.g.}, Ref.~\cite{2015PhRvD..92f3009K}). However, it is not drastically dissimilar to recent calculations using the BSk family of equations of state, where a \textit{g}-mode solution was found with frequency in excess of \qty{250}{\hertz} for BSk22 \cite{2025MNRAS.536.1967C}. The high frequency is due to the large compositional gradients supported by the APR equation of state, which is shown in the right panel of Fig.~\ref{fig:Profiles}. Around neutron drip, where the \textit{g}\textsubscript{2}-mode eigenfunction changes behaviour (see Fig.~\ref{fig:g-modes}), the APR equation of state at $T = \qty{0.02}{\mega\electronvolt}$ has larger composition gradients, which provide a greater buoyancy force for oscillations in this region.

The left panel of Fig.~\ref{fig:Spectrum} shows the spectrum of the rest of the \textit{g}-modes. We see that these oscillations are consistent with typical \textit{g}-mode frequencies of $\text{Re}(\omega) / (2 \pi) \sim \qtyrange{10}{100}{\hertz}$ \cite{1992ApJ...395..240R,1994MNRAS.270..611L,2015PhRvD..92f3009K}. We summarise the oscillations of the cold neutron star in Table~\ref{tab:Cold}.

\begin{table}[ht]
    \caption{\label{tab:Cold}%
    The (real) eigenfrequencies of the quadrupolar \textit{f}-mode and first several \textit{g}-modes of the cold neutron star described by the APR equation of state.}
\begin{tabular}{ l @{\qquad} S }
    \toprule
     & \protect{$\text{Re}(\omega) / (2 \pi)$ / \unit{\hertz}} \\
    \colrule
    \textit{f}-mode & 1912.3 \\
    \textit{g}\textsubscript{1}-mode & 508.5 \\
    \textit{g}\textsubscript{2}-mode & 304.5 \\
    \textit{g}\textsubscript{3}-mode & 126.1 \\
    \textit{g}\textsubscript{4}-mode & 116.6 \\
    \textit{g}\textsubscript{5}-mode & 112.6 \\
    \textit{g}\textsubscript{6}-mode & 89.1 \\
    \textit{g}\textsubscript{7}-mode & 72.1 \\
    \textit{g}\textsubscript{8}-mode & 71.7 \\
    \textit{g}\textsubscript{9}-mode & 61.8 \\
    \botrule
\end{tabular}
\end{table}

\subsection{\label{sec:Temp}Thermal treatments}

\begin{figure}
    \includegraphics[width=0.5\textwidth]{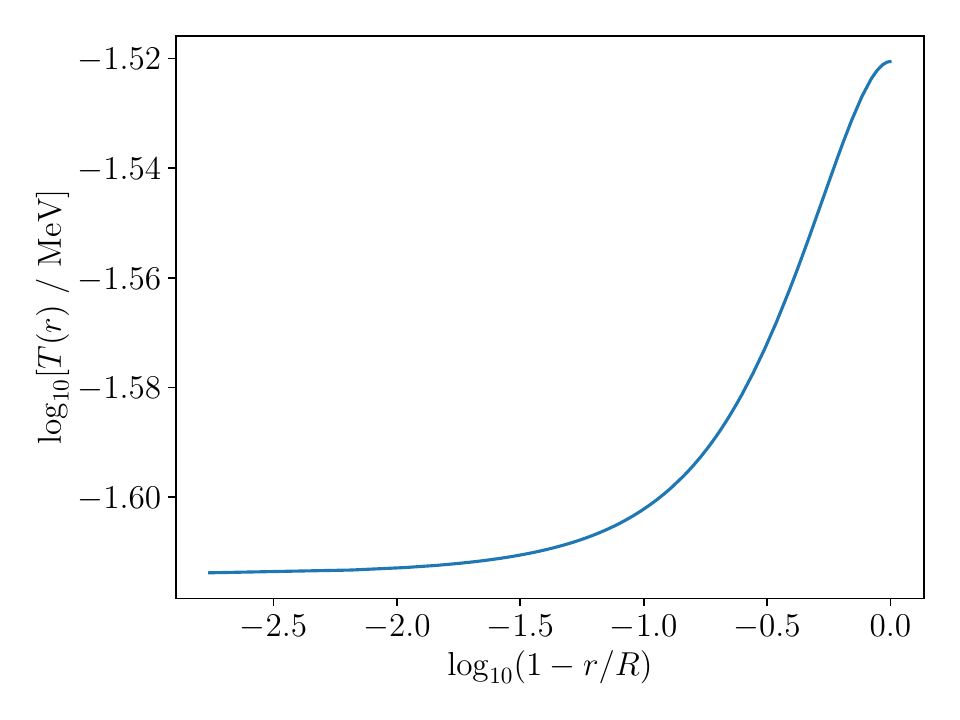}%
    \includegraphics[width=0.5\textwidth]{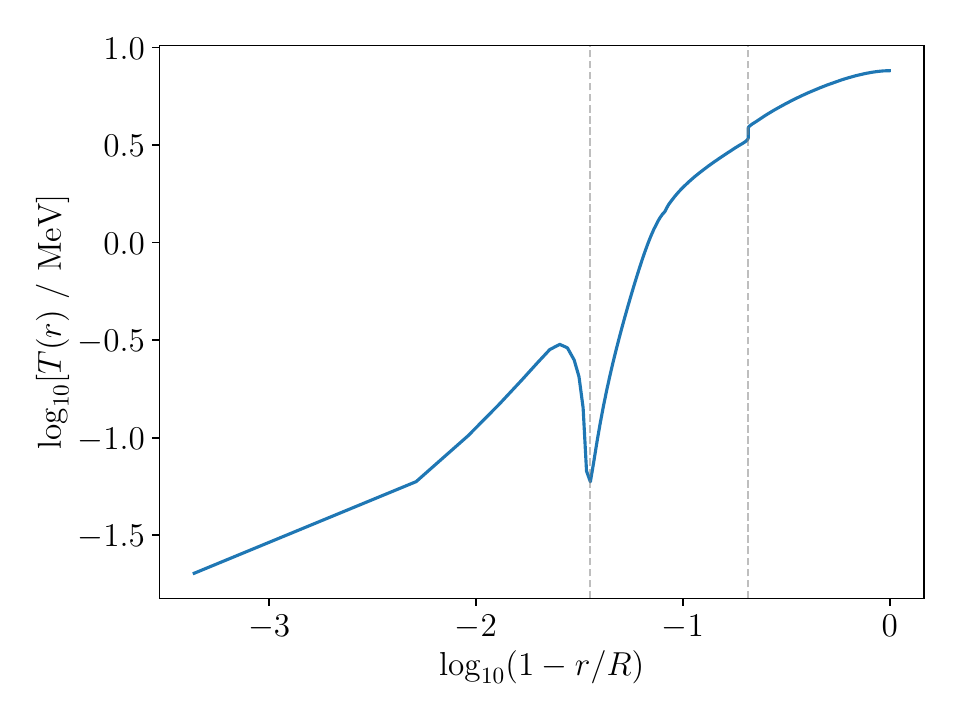}
    \caption{\label{fig:Temp}%
    The temperature profile of a neutron star with constant redshifted temperature of $T e^{\nu / 2} = \qty{0.02}{\mega\electronvolt}$ (left panel) and uniform entropy per baryon of $s = 0.25$ (right panel). In the right panel, neutron drip (left) and the APR phase transition (right) are indicated with vertical, dashed lines. These two aspects of the matter model give rise to sharp features in the temperature distribution. Both stars have similar temperatures at the surface, $T \sim \qty{0.02}{\mega\electronvolt}$, whereas the differences are appreciable within the interior.}
\end{figure}

Having laid foundations in the cold regime, we now turn our attention to exploring how the spectrum is modified with higher temperatures. A typical neutron star is expected to be formed from the remnant of a supernova. The neutron star will be born hot and rapidly rotating, and will cool due to the emission of photons and neutrinos \cite{1986ApJ...307..178B,1994ApJ...425..802L,2001MNRAS.324..725G}. The expectation in general relativity is that the star will eventually evolve towards uniform redshifted temperature, where $T e^{\nu / 2} = \text{const}$ \cite{1967hea3.conf..259T}. However, in the intermediate phases during the cooling, the neutron star will host a non-trivial temperature profile. Indeed, recent studies of the thermal evolution show that the matter reaches uniform entropy per baryon, $s = \text{const}$, in the core quite rapidly \cite{2022MNRAS.511..356P} and that the star obtains constant redshifted temperature later \cite{2023ApJ...942...72B}. Therefore, in this section, we will explore two temperature prescriptions for the neutron star: uniform redshifted temperature and constant entropy per baryon. In Fig.~\ref{fig:Temp}, we present representative examples of these two cases. In our integrations, the surface of the neutron star was defined by when the pressure reaches the smallest value for the given local temperature in the equation-of-state table.

We examine uniform redshifted temperature first. In order to compute the stellar equilibrium, we now specify the redshifted temperature and use the shooting method to satisfy the boundary condition on the metric potential at the surface, $e^{\nu(R)} = 1 - 2 M / R$. The star's temperature profile results in a non-zero $dT / dn_\text{b}$ (see the left panel of Fig.~\ref{fig:Temp}). To determine this quantity, we first note that
\begin{equation}
    \frac{dT}{dn_\text{b}} = - \frac{T}{2} \frac{d\nu}{dn_\text{b}} = \frac{T}{\varepsilon + p} \frac{dp}{dn_\text{b}}.
\end{equation}
Then, combining Eqs.~\eqref{eq:dp_dnb} and \eqref{eq:dYe_dnb}, we obtain
\begin{equation}
    \frac{dT}{dn_\text{b}} = \left[ \left( \frac{\partial p}{\partial n_\text{b}} \right)_{T, Y_\text{e}} + \left( \frac{\partial p}{\partial Y_\text{e}} \right)_{T, n_\text{b}} \left( \frac{\partial Y_\text{e}}{\partial n_\text{b}} \right)_T \right] \left[ \frac{\varepsilon + p}{T} - \left( \frac{\partial p}{\partial T} \right)_{n_\text{b}, Y_\text{e}} - \left( \frac{\partial p}{\partial Y_\text{e}} \right)_{T, n_\text{b}} \left( \frac{\partial Y_\text{e}}{\partial T} \right)_{n_\text{b}} \right]^{-1}.
\end{equation}
This information is required to determine $\Gamma$.

As a consistency check, we calculated the perturbations of a neutron star with $T e^{\nu / 2} = \qty{0.02}{\mega\electronvolt}$ and found very few differences as compared to the cold constant-temperature model. The eigenfrequencies as reported in Table~\ref{tab:Cold} are the same to the accuracy provided.

\begin{figure}
    \includegraphics[width=0.5\textwidth]{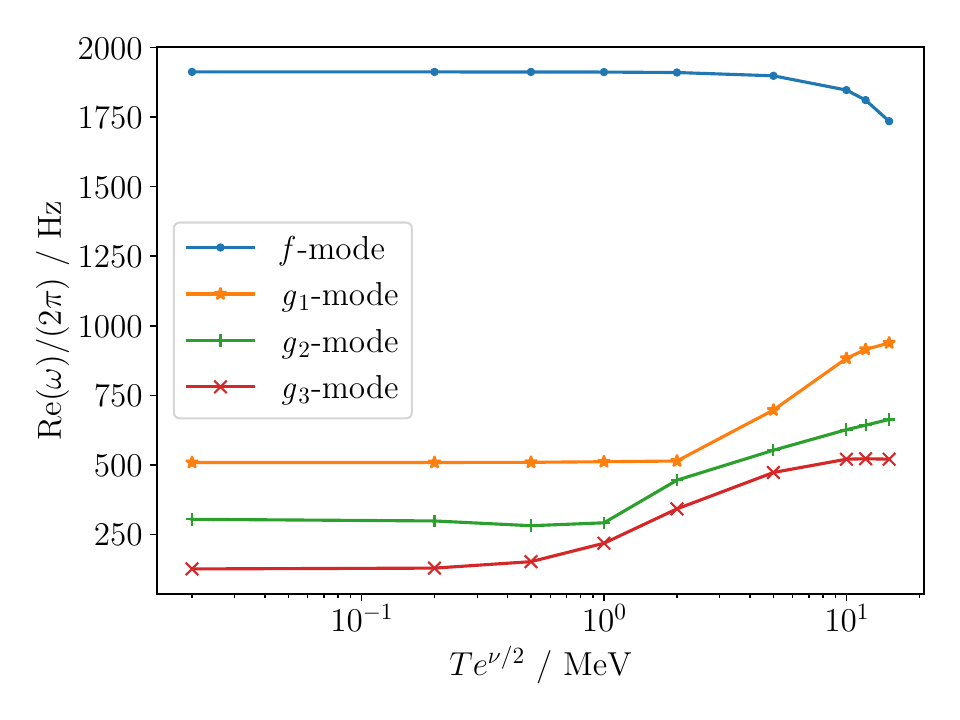}%
    \includegraphics[width=0.5\textwidth]{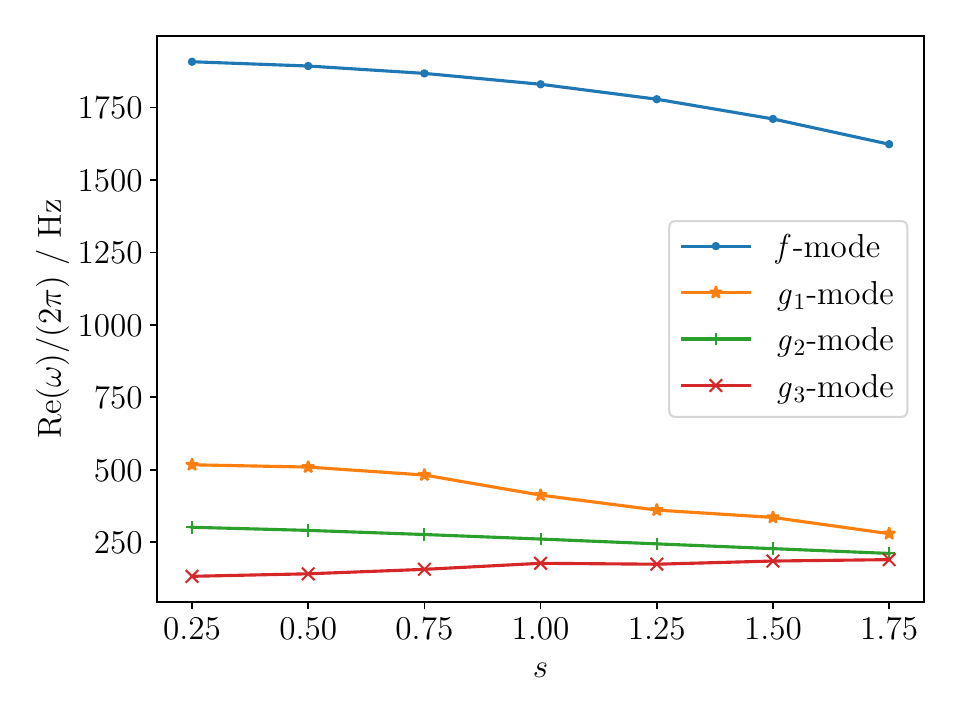}
    \caption{\label{fig:Frequencies}%
    The (real) quadrupolar mode eigenfrequencies of the neutron star against redshifted temperature (left panel) and entropy per baryon (right panel). The eigenfrequencies correspond to the \textit{f}- and first three \textit{g}-modes. The \textit{f}-mode is relatively insensitive to the temperature in both cases, even up to temperatures of $T \sim \qty{1}{\mega\electronvolt}$. As the star heats up, in the two panels, the \textit{f}-mode gradually decreases in frequency. This is due to the expansion in stellar radius. In contrast, the \textit{g}-modes change drastically as the higher redshifted temperatures source larger entropy gradients. However, in the second case, there are no entropy gradients and the \textit{g}-modes are supported entirely by composition gradients. The \textit{g}\textsubscript{1}- and \textit{g}\textsubscript{2}-modes steadily decrease in frequency, whereas the \textit{g}\textsubscript{3}-mode rises.}
\end{figure}

Moving on, we now examine higher temperatures of the neutron star. We determined the frequencies of the \textit{f}-mode and first three \textit{g}-modes as functions of the redshifted temperature in the range $T e^{\nu / 2} = \qtyrange{0.02}{15}{\mega\electronvolt}$. This sequence of neutron stars share the same the baryon mass of $M_\text{b} = 1.4 M_\odot$. As the star is heated up, it goes from an equilibrium radius of $R = \qty{11.6}{\kilo\metre}$ up to $R = \qty{16.6}{\kilo\metre}$. The oscillation frequencies are displayed in the left panel of Fig.~\ref{fig:Frequencies}.

As expected, the frequency of the \textit{f}-mode---approximately proportional to $\sqrt{M / R^3}$---decreases as the star expands. Figure~\ref{fig:Frequencies} shows that the \textit{f}-mode frequency is relatively insensitive to the temperature up to $T e^{\nu / 2} \sim \qty{1}{\mega\electronvolt}$. The situation is quite different for the \textit{g}-modes.

\begin{figure}
    \includegraphics[width=0.5\textwidth]{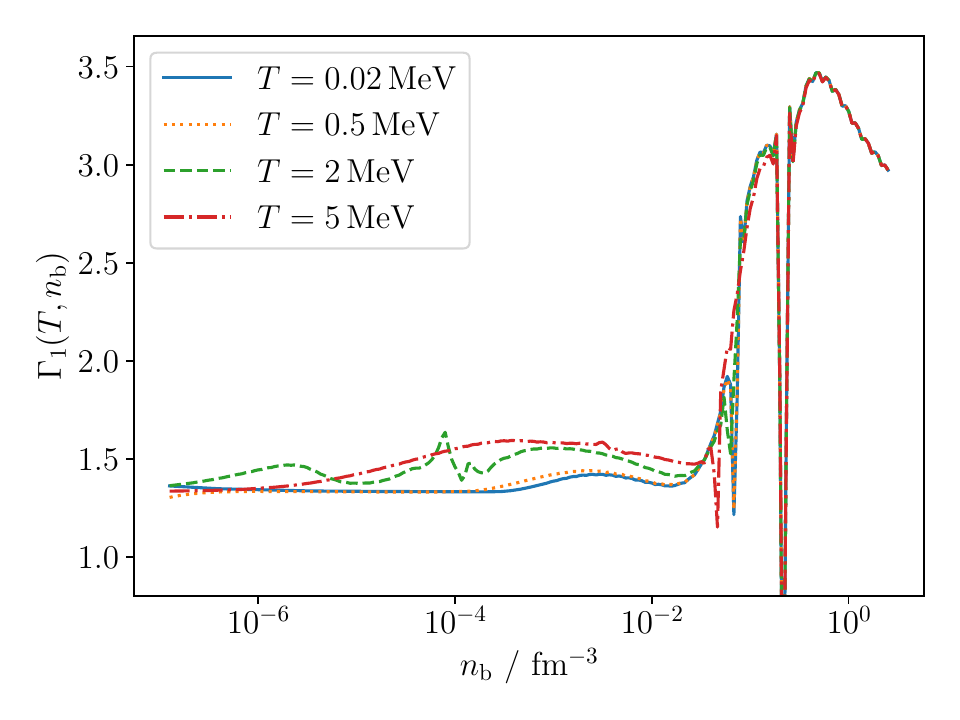}
    \caption{\label{fig:Gamma_1}%
    The adiabatic index against baryon-number density for finite-temperature APR nuclear matter. Each line corresponds to a different (uniform) temperature, as indicated by the legend. As the temperature increases, the matter is able to support enhanced entropy gradients leading to higher frequency \textit{g}-modes.}
\end{figure}

Thermodynamically, temperature and entropy are conjugates of each other. Thus, as the star is heated, the entropy gradients in the fluid become more pronounced. We illustrate this effect in Fig.~\ref{fig:Gamma_1}, where we show how $\Gamma_1$ enlarges at lower densities with increasing temperature. The steeper entropy gradients provide a stronger restoring force for buoyant perturbations and cause the \textit{g}-modes to oscillate more rapidly. This is consistent with the behaviour seen in Fig.~\ref{fig:Frequencies}. These results are in qualitative agreement with other studies \cite{2003MNRAS.342..629F,2015PhRvD..92f3009K}.

Next, we consider the case of uniform entropy per baryon throughout the star. To generate the background, we inverted the equation-of-state table to make $s$ an independent variable, rather than $T$. From this information, it is trivial to pick a fixed $s$ and solve the structure equations. We display the resultant temperature profile for $s = 0.25$ in the right panel of Fig.~\ref{fig:Temp}. We note that, in this case, the profile is sensitive to neutron drip and the phase transition in the equation-of-state model.

To calculate the oscillation modes, we interpolated the temperature as $T = T(s, n_\text{b})$, assuming chemical equilibrium, and obtained its derivative $dT / dn_\text{b}$ at fixed entropy per baryon from the splines. This was sufficient information to then compute the adiabatic index $\Gamma$. We explored the range of $s = \numrange{0.25}{1.75}$ and the oscillation frequencies are shown in the right panel of Fig.~\ref{fig:Frequencies}. The background stars have $M_\text{b} = 1.4 M_\odot$ and the radii vary from $R = \qty{11.6}{\kilo\metre}$ to $R = \qty{14.1}{\kilo\metre}$.

The dependence of the \textit{f}-mode on the entropy per baryon is similar to the constant-redshifted-temperature case. As the star expands, its frequency decreases. However, the \textit{g}-modes have a different behaviour. Since there are no entropy gradients for this case, the \textit{g}-mode oscillations are supported by variations in the chemical composition. It can be seen in Fig.~\ref{fig:Frequencies} that the first two \textit{g}-modes steadily decrease their frequencies with increasing entropy per baryon. This is in notable contrast to the previous thermal model. The third \textit{g}-mode frequency, however, increases.

\section{\label{sec:Conc}Conclusions}

In this paper, we have calculated the oscillation spectra of neutron stars subject to finite-temperature nuclear matter, implementing a realistic, tabulated equation of state. We discussed how realistic, finite-temperature nuclear matter enters the computation of neutron-star oscillation modes. In paying close attention to the thermodynamics, we have illustrated the implicit assumptions with one-parameter (barotropic) matter models; zero temperature $T = 0$ and the particle species maintaining chemical equilibrium $\mu_\Delta = 0$. These assumptions will not hold for dynamical situations, such as oscillating neutron stars, proto-neutron stars and during compact-binary coalescences.

With a view to constructing realistic neutron-star models, we have focused our attention on three-parameter equations of state, which represents the present state of the art in nuclear astrophysics. With these equations of state, the stellar material can support entropy and composition gradients, giving rise to low-frequency \textit{g}-mode oscillations.

As a demonstration, we adopted the APR equation of state with finite temperature and examined the resultant oscillation spectrum for $T = \qty{0.02}{\mega\electronvolt}$. APR nuclear matter involves a discontinuity in the energy density that corresponds with the phase transition between nucleonic matter and a neutral pion condensate. We identified a discontinuity \textit{g}-mode sourced by this interface in the star. The likely explanation for its high frequency $\text{Re}(\omega) / (2 \pi) = \qty{508.5}{\hertz}$ is the large density jump in the APR equation of state. We also calculated the first few \textit{g}-modes that arise from composition and entropy stratification in the fluid.

Realistic, mature neutron stars are expected to possess non-trivial temperature profiles as they gradually cool. We examined the oscillation spectrum under two simple prescriptions for the temperature dependence of the star. In the first case, we considered uniform redshifted temperature and showed how the eigenfrequencies of the \textit{g}-modes increased substantially with larger entropy gradients. For the second prescription, we immersed the star in a constant entropy-per-baryon profile. We noted how the first two \textit{g}-modes gradually decreased in frequency with varying composition gradients, whereas the third \textit{g}-mode frequency steadily increased.

As we wrap up, it is worth making some final comments on the implementation of the thermodynamics. Matter models are typically provided as tables, owing to the numerical complexity of computing nuclear interactions at high densities. Since neutron-star models require this information in functional form, we must resort to numerical interpolation of the tabular data. We chose to work with splines, since they are computationally efficient to use and partial derivatives can be obtained trivially. A natural improvement to our approach would be to use an interpolation method that is thermodynamically consistent \cite{1982ApJ...258..306H,1996JCoPh.127..118S}. Indeed, the procedure outlined in Ref.~\cite{1996JCoPh.127..118S} is implemented in popular, open-source neutron-star codes {\footnotesize RNS} \cite{rns} and {\footnotesize LORENE} \cite{lorene}. We leave this extension for future work.

Additionally, the oscillation modes depend on thermodynamic derivatives---such as the first partial derivatives of $p = p(T, n_\text{b}, Y_\text{e})$. Even in the cases where this information is provided in the equation-of-state table, it is likely obtained numerically, since the models do not lend themselves to analytical forms. When this is the case, numerical differentiation is the only option. We determined this information by interpolating the quantities using splines and used the spline coefficients to estimate the derivatives. As can be seen in Fig.~\ref{fig:Profiles}, this gives rise to singular features in the neutron-star profile, which leave their mark on the oscillation spectrum. Along this vein, there have been some welcome attempts to generate analytical representations for thermal \cite{2019ApJ...875...12R,2024PhRvC.109e5202L} and compositional features \cite{2024PhRvD.109j3022S}. With the goal of conducting reliable neutron-star asteroseismology in the era of precision astronomy, these are important issues to address.

\begin{acknowledgements}
   F.G. acknowledges funding from the European Union’s Horizon Europe research and innovation programme under the Marie Sk{\l}odowska-Curie grant agreement No.~101151301.
   N.A. acknowledges support from STFC via grant No.~ST/R00045X/1.
   The authors are grateful to Ian Hawke for useful discussions on the numerical component of this work and Christian Kr{\"u}ger for sharing code to compare against. The software developed to support this article is available in a GitHub repository \cite{code} and is written in the {\footnotesize Julia} programming language \cite{Bezanson:2014pyv,DifferentialEquations.jl-2017,Interpolations.jl,2024arXiv240316341P,Optim.jl-2018}. The figures were generated using {\footnotesize Matplotlib} \cite{2007CSE.....9...90H,PythonCall.jl}.
\end{acknowledgements}

\bibliography{refs.bib}

\end{document}